\documentclass{hep99}
\input psfig.sty
\def\etal{{\sl et al.\/}\ }                 %et al. - no preceeding comma

\begin{document}
\title{Search for Exotics at the Tevatron: Present and Future}
\author{Greg Landsberg\\(on behalf of the CDF and D\O\ Collaborations)}
\address{Brown University, Dept. of Physics, 182 Hope St, Providence, RI 02912}
\abstract{
We present recent results on searches for non-SUSY new physics at the CDF and D\O\ Collaborations in the 1992--1996 Fermilab Tevatron run. While no compelling evidence for existence of new physics was found, the Tevatron data have excluded a significant region of the theoretically allowed phase space for a variety of non-SUSY extensions of the Standard Model. Tight limits on the existence of the following new phenomena are set: leptoquarks of all  three generations, quark-lepton compositeness, 4-th generation quarks, fermiophobic Higgs, technicolor, etc. Prospects of the Tevatron experiments in Run II are discussed.}
\maketitle

\section{Introduction}

While SUSY remains a very interesting theoretical possibility, the quest for new physics at the Fermilab Tevatron is not restricted to this class of models. Following the classical ``presumption of innocence'' in science, both the CDF and D\O\ collaborations have performed extensive tests of a variety of new phenomena predicted in various extensions of the Standard Model (SM). The results presented in this talk are just a sample of this extensive search program, and cover only the most recently analyzed signatures.

Details of the following analyses are presented: search for all three generations of  leptoquarks, an attempt to find quark-lepton and quark-quark compositeness, a quest for extra generations of matter, fermiophobic Higgs boson, technicolor.

\section{Search for Leptoquarks}

\begin{figure*}
\centerline{\hbox{
\protect\psfig{figure=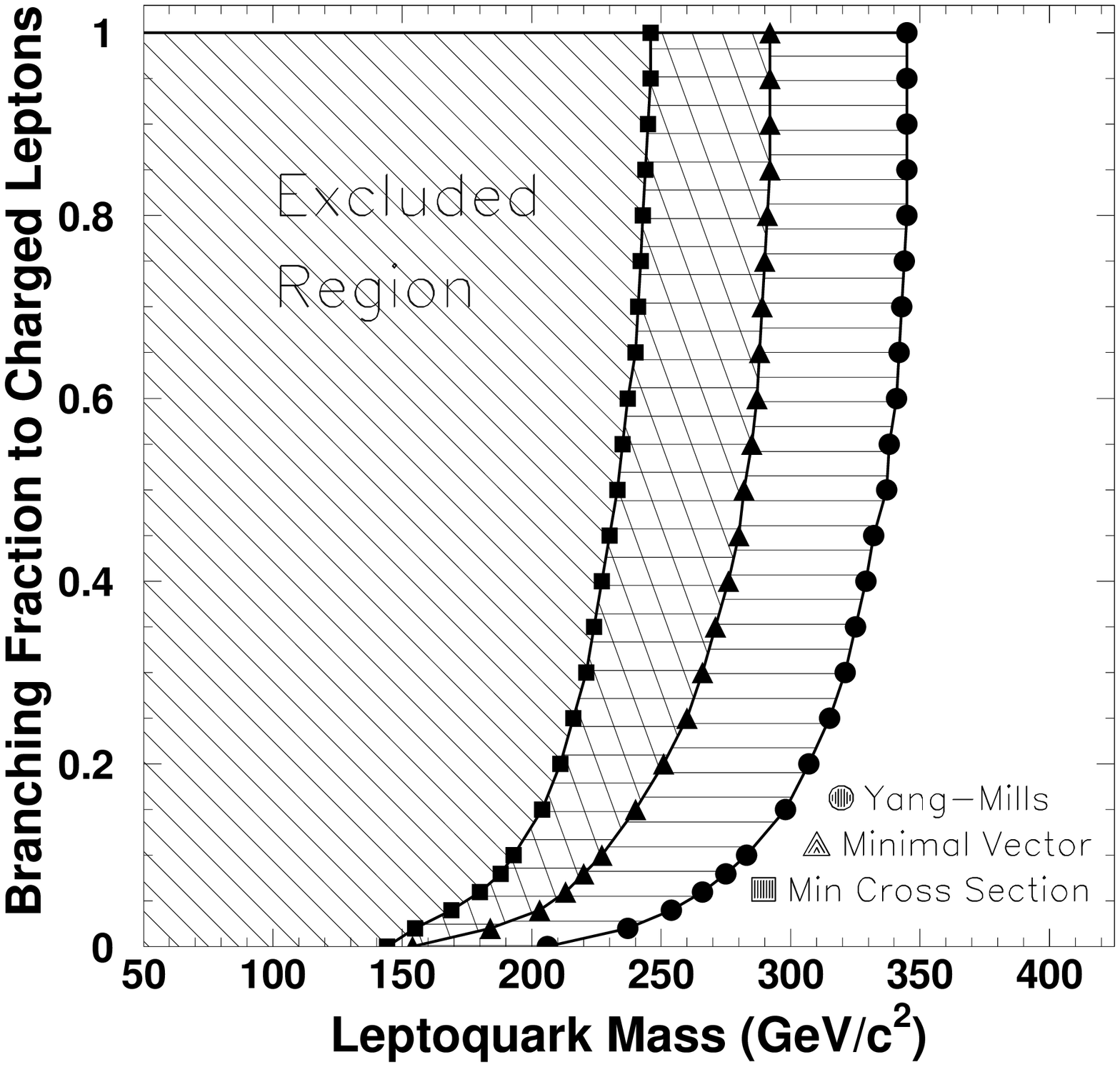,width=2.9in}
\protect\psfig{figure=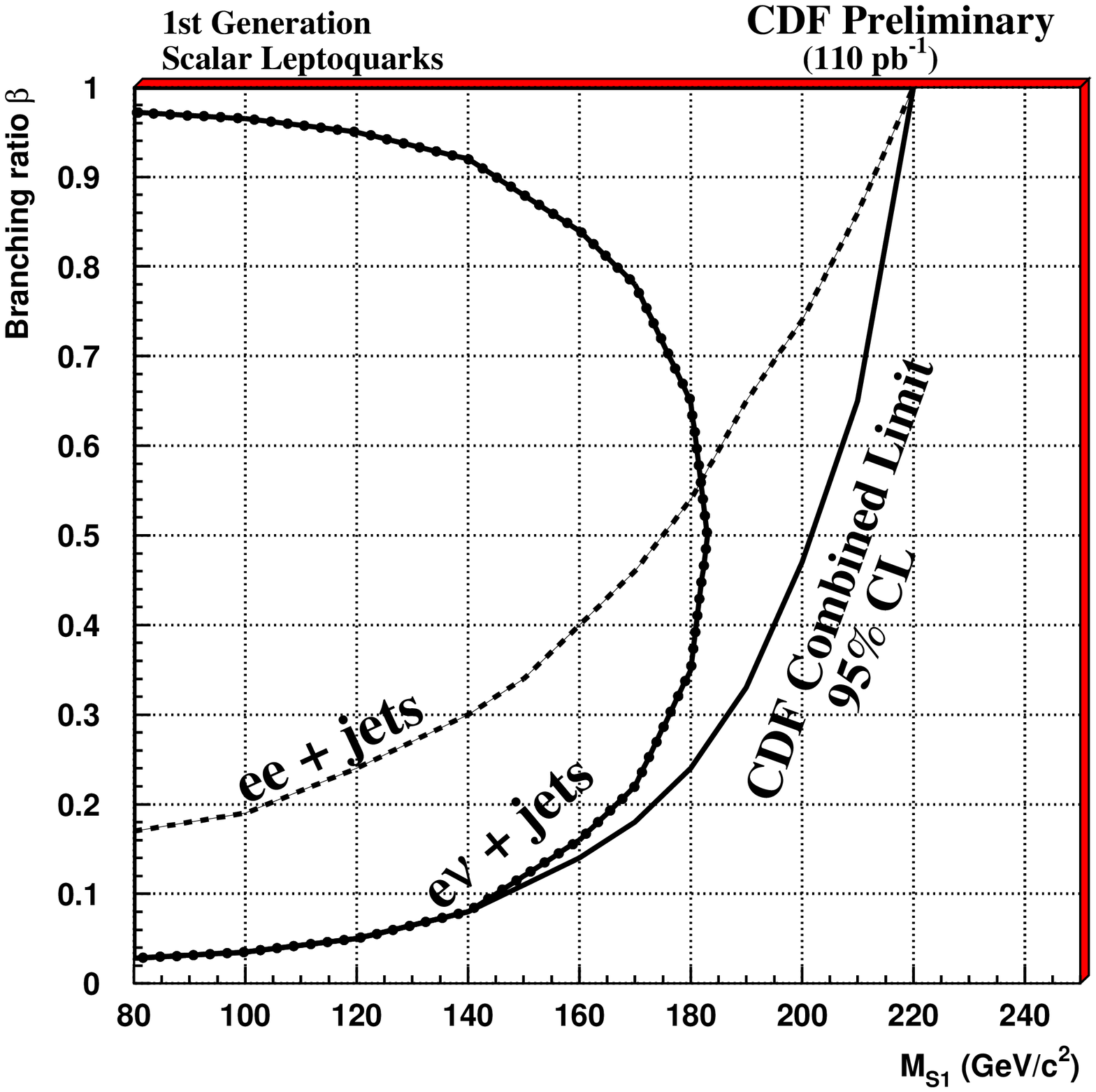,width=2.9in}
}}
\caption{a) D\O\ 95\% C.L. lower limits on the vector LQ1 mass for three different values of vector couplings; b) CDF preliminary 95\% C.L. lower limits on the scalar LQ1 mass as a function of branching fraction into charged lepton mode $\beta$ from the $ee+{\rm jets}$ and $e\nu+{\rm jets}$ channels.}
\label{fig:LQ1}
\end{figure*}

Leptoquarks~\cite{Pati-Salam} (LQ) are hypothetical objects that carry properties of both leptons and quarks and appear as a natural extension of the SM in many GUT-inspired theories. Recent interest in the LQ was inspired by an observation of a high-$Q^2$ event excess in the $e^+ p$-collisions at HERA~\cite{LQ-boom}. Both the D\O\ and CDF Collaborations have published the results of their searches for the LQ of the first generation (LQ1) that excluded an explanation of HERA event excess by the LQ1 production~\cite{LQ1-TeV}. D\O\ has analyzed all three possible signatures for LQ1 pair production ($ee+{\rm jets}$, $e\nu + {\rm jets}$, and $\nu\nu + {\rm jets}$), while CDF has published results in the dielectron mode. Recently, D\O\ has completed their LQ1 analysis by producing limits on the vector leptoquarks of the first generation as a function of the LQ branching ratio in the charged lepton channel, $\beta$. These limits for the case of different vector couplings are shown in Fig.~\ref{fig:LQ1}a. The CDF collaboration has recently obtained preliminary results using a more optimized LQ1 analysis strategy in both $ee + {\rm jets}$ and $e\nu + {\rm jets}$ channels, as shown in Fig.~\ref{fig:LQ1}b. These new limits are similar to those from the published D\O\ analyses.

New H1 limits on the LQ1~\cite{H1} are comparable with those from the Tevatron for large values of coupling between the LQ's, leptons, and quarks, and are complementary in nature.

\begin{figure}
\centerline{
\protect\psfig{figure=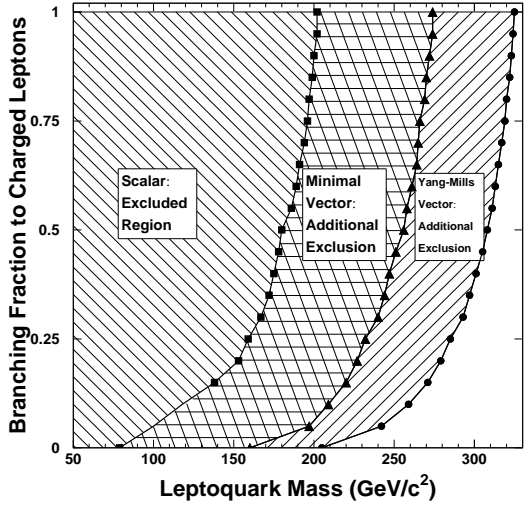,width=2.7in}
}
\caption{D\O\ preliminary 95\% C.L. lower limits on the scalar  and vector LQ2 mass for two different values of vector couplings.}
\label{fig:LQ2}
\end{figure}

Both Tevatron Collaborations have also searched for the second (LQ2) and third (LQ3) generation leptoquarks. CDF has recently published a lower 95\% C.L. LQ2 mass limit of 202 GeV (160 GeV) for $\beta = 1$ ($\beta = 1/2$) based on the $\mu\mu + {\rm jets}$ channel~\cite{LQ2-CDF}. CDF has also obtained a preliminary lower 95\% C.L. LQ2 mass limit of 122 GeV for $\beta = 0$ in the $c\nu$ decay mode of LQ2. D\O\ has recently completed the LQ2 searches in all three modes~\cite{LQ2-D0} and the corresponding limits as a function of $\beta$ for scalar and two vector LQ couplings are shown in Fig.~\ref{fig:LQ2}. While the D\O\ sensitivity for $\beta = 1$ is the same as that for CDF, D\O\ limits at $\beta = 1/2$ are more restrictive, due to the addition of the $\mu\nu + {\rm jets}$ channel. The D\O\ limit for $\beta = 0$ based on the flavor-independent $\nu\nu + {\rm jets}$ search is less restrictive than that from the dedicated CDF search that benefited from the $c$-jet tagging.

\begin{figure}
\centerline{
\protect\psfig{figure=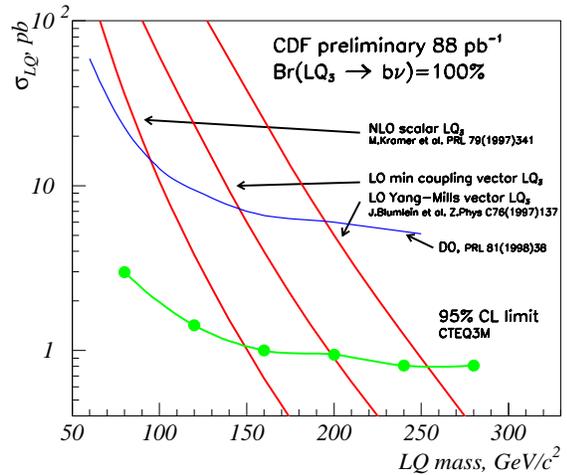,width=3.2in}
}
\caption{CDF preliminary 95\% C.L. lower limits on the scalar LQ3 mass from the $b\nu$ decay channel.}
\label{fig:LQ3}
\end{figure}

CDF has obtained recently preliminary results of a LQ3 search in the $\nu\nu + b\bar b$ channel with one or more $b$-tagged jets required, a significant improvement over the previously published D\O\ limits~\cite{LQ3-D0}. This result (see Fig.~\ref{fig:LQ3}) complements a previously published CDF analysis in the $\tau$-channel~\cite{LQ3-CDF}.

\begin{figure*}[htb]
\centerline{\hbox{
\protect\psfig{figure=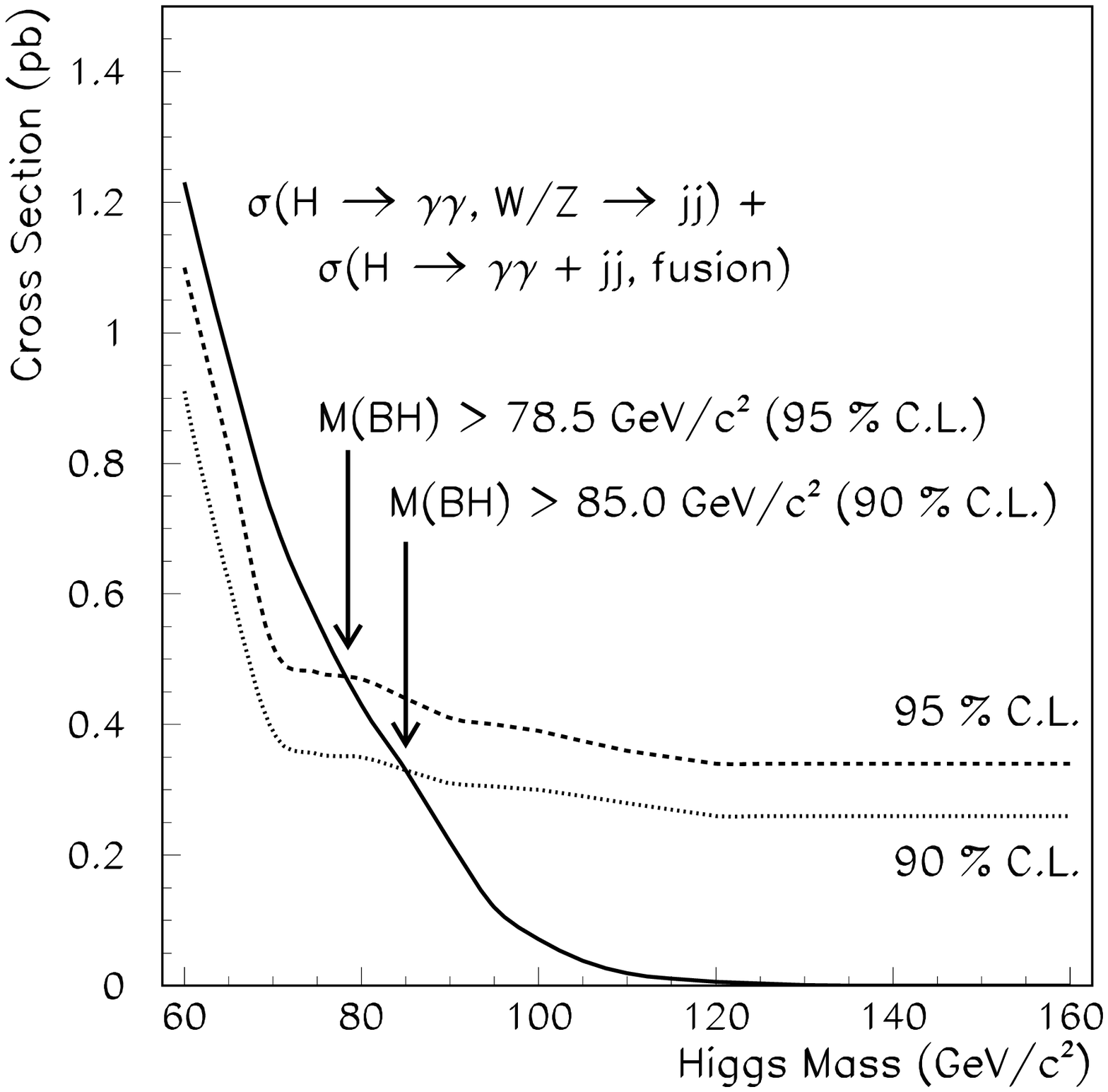,width=2.8in}
\protect\psfig{figure=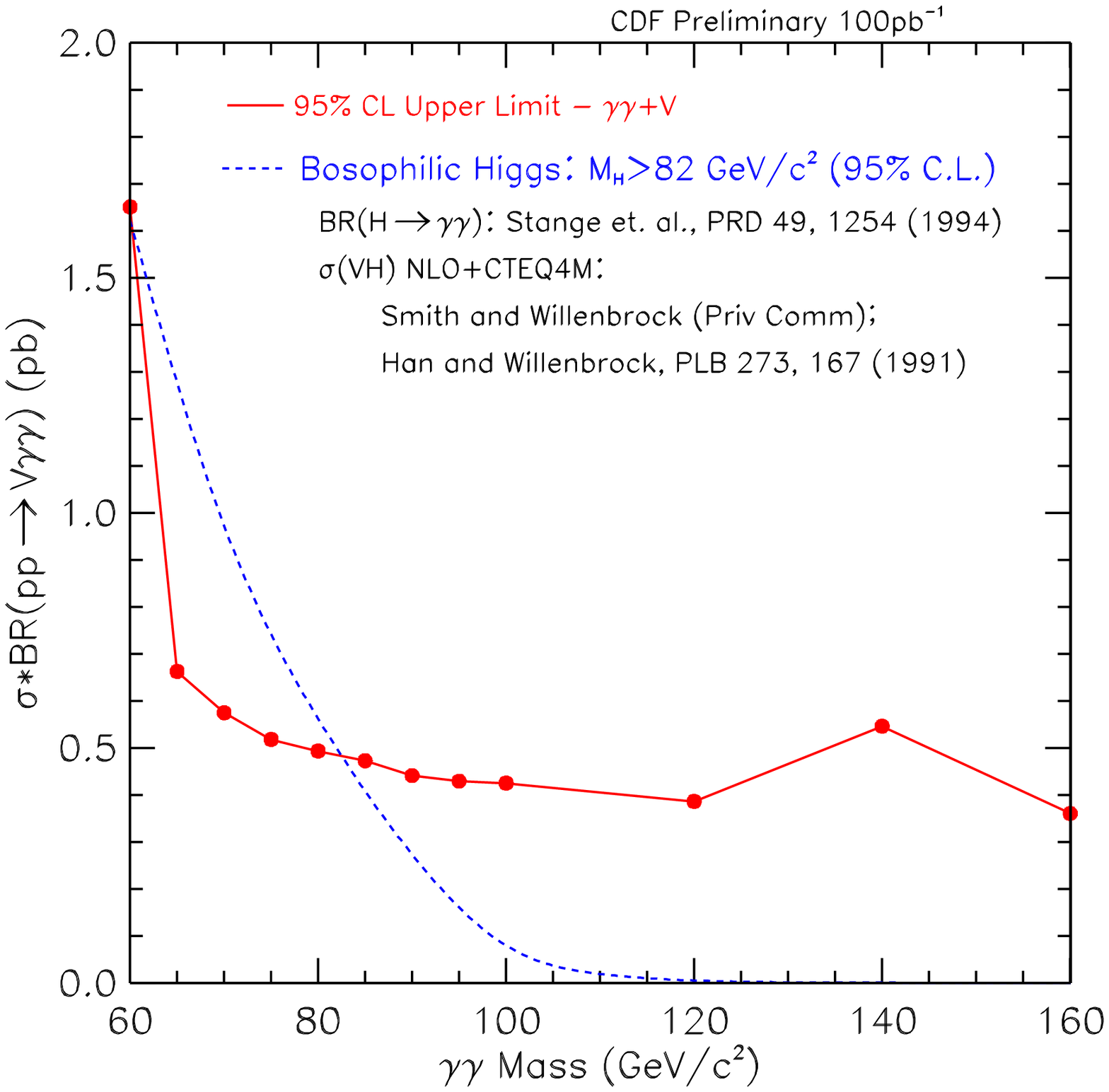,width=2.6in}
}}
\caption{a) D\O\ 90\% and 95\% C.L. and b) CDF preliminary 95\% C.L. lower limits on the mass of a fermiophobic Higgs boson.}
\label{fig:Higgs}
\end{figure*}

\section{Search for Quark and Lepton Compositeness}

Since ancient times the human mind has tried to identify the smallest constituents of the Universe. Currently, we think that quarks and leptons are such indivisible building blocks of  matter. However, in certain models, quarks and leptons are also composite particles that consist of even smaller species~-- preons. If quarks or leptons are composite, the Drell-Yan  Lagrangian is modified with an additional, contact term that dominates at the energies comparable with the compositeness scale (or the inverse size of the above mentioned substructure). Therefore, Drell-Yan production of pairs of quarks or leptons allows us to probe possible composite structures of these particles. Both the CDF and D\O\ Collaborations looked for compositeness in various channels. CDF has published a search for compositeness in dielectron and dimuon production~\cite{CDF-composite}, while D\O\ has looked in the dijet and dielectron channels~\cite{D0-composite}. Neither of the Collaborations found any deviations from the theory based on pointlike quarks and leptons, and set limits on the compositeness scale between 2.1 and 6.1 TeV, depending on the parameters of the compositeness model. Recent LEP~II limits typically exceed the sensitivity of the Tevatron experiments, but in many cases they are complementary to the CDF/D\O\ measurements due to a different production mechanism. For example, $q\bar q \to q \bar q$ production mechanism is unique to the Tevatron, while $e^+ e^- \to \ell^+\ell^-$ is unique to LEP~II.

\section{Search for Extra Generations of Matter}

While the number of light neutrino species, and hence lepton generations, is precisely known to be three from the $Z$-peak data, it is much harder to constrain the number of quark generations. Not surprisingly, both Tevatron Collaborations have looked for pair production of the fourth-generation charge $-1/3$ quark, $b'$. D\O\ has published~\cite{b'-D0} a search in the mode where the $b'$-quark decays into $\gamma b$ or $gb$ modes, expected to be dominant for a mass of the $b'$-quark below the kinematic threshold of the competing, $Zb$ channel (the latter is expected to dominate for $M(b') > 100$~GeV). The results of this search have excluded light $b'$-quark with masses up to the kinematic limit, 95 GeV. Recently, CDF has obtained preliminary results on the $b'$-quark decaying into the $Zb$ channel~\cite{b'-CDF}. Assuming $b'$ pair production with one of the daughter $Z$-bosons decaying hadronically, and the other leptonically, that corresponds to a $\ell^+\ell^- + {\rm jets}$ channel. One or more jets in this analysis were required to be $b$-tagged using the microvertex detector. The search has not revealed any excess of data above the expected SM background and thus excluded $b'$ masses between 100 and 175~GeV under the assumption of 100\% branching fraction of the $b' \to Zb$ process. Alternatively, this result can be interpreted as the upper limit on the branching fraction as a function of the $b'$ mass (branching fractions above 40--75\% are excluded at 95\% C.L.). Between the CDF and D\O, virtually all $b'$ masses below 175~GeV are now excluded, assuming that the FCNC decay mode is dominant. This is yet another example of complementarity of results from the two Tevatron Collaborations.

\section{Search for Fermiophobic Higgs}

In certain extensions of the SM, the Higgs particle is expected to have an enhanced decay into photons. Both the CDF and D\O\ Collaborations studied the case where Higgs decay to fermions is prohibited~\cite{Higgs-theory}, and therefore the $\gamma\gamma$ decay mode is dominant for light Higgs (below the $W^*W$ threshold). At the Tevatron, Higgs is dominantly produced in association with a $W$ or $Z$ boson, so the final state is $\gamma\gamma + 2$~jets. D\O\ has recently published~\cite{Higgs-D0} an analysis in that mode that resulted in a 95\% C.L. lower limit on fermiophobic Higgs mass of 78.5~GeV (see Fig.~\ref{fig:Higgs}a). The analogous preliminary analysis from CDF has a similar limit of 82~GeV (see Fig.~\ref{fig:Higgs}b). Recent preliminary results from LEP~II experiments (see these proceedings) raised the fermiophobic Higgs mass limit to $\approx 96$ GeV.

\section{Search for Technicolor}

CDF Collaboration has recently published a set of new limits on technicolor (TC) and topcolor models~\cite{TC-CDF}. These results include a search for $q\bar q' \to \rho_T \to W\pi_T$ in lepton + jets and four-jet modes, search for $q\bar q \to \omega_T \to \gamma\pi_T$, search for color-octet $\rho_T$ and topgluons in the $b\bar b$ mode, and search for enhanced production of third generation leptoquarks via $q \bar q' \to \rho_T \to {\rm LQ3} + {\rm LQ3}$ modes. These pioneer searches for TC at hadron colliders resulted in stringent limits on the existence of technicolor particles. In particular, a color octet $\rho_T$ with mass less than about 650~GeV has been excluded at 95\% C.L., as well as significant regions in the color singlet $\rho_T-\pi_T$ mass plane.

\section{Conclusions}

This is just a sample of new physics searches recently carried out by the CDF and D\O\ Collaborations. Both Tevatron experiments have looked for a broad variety of different non-SUSY new physics phenomena, including additional gauge bosons, excited quarks and fermions. With many analyses still under way, we expect a steady flow of new physics results to be available from both Collaborations until the start of the next Tevatron run in early 2001. With the enhanced detectors, higher collider energy, as well as much higher integrated luminosity, we expect an outstanding search potential in the next Run and we are looking forward to start it soon. We expect the sensitivity to pair-produced objects in terms of their mass to increase by about 100 GeV; even larger improvement is expected for single or associated production of new particles. The reader could either stay tuned or try to join us in this fun, as many European groups have done recently. After all, doing Tevatron physics is the best way to prepare to do physics at the LHC.

\end{document}